\documentclass[twocolumn]{article}
\usepackage{fullpage}

\usepackage{cite}

\usepackage{amsthm,amsmath,amssymb,amsfonts}
\usepackage{graphicx}
\usepackage{textcomp}

 \usepackage{hyperref}
\usepackage{flushend}
 \usepackage[table,xcdraw]{xcolor}
\usepackage[linesnumbered,algoruled,boxed,lined]{algorithm2e}

\newcommand{\ie}{\emph{i.e.}, }
\newcommand{\eg}{\emph{e.g.}, }
\newcommand{\M}{\mathcal{M}}

\newcommand{\sm}{{\tt SM}\xspace}
\newcommand{\wg}{{\tt WGHT}\xspace}
\newcommand{\gr}{{\tt GRID}\xspace}
\newcommand{\adv}{{\tt BADV}\xspace}

\DeclareMathOperator*{\argmax}{arg\,max}

\usepackage[makeroom]{cancel}

\usepackage{subfigure}

\newtheorem{definition}{Definition}
\theoremstyle{plain}

\newtheorem{problem}{Problem}

\usepackage{ color, colortbl}
\definecolor{Gray}{gray}{0.9}

\begin{document}

\title{TamperNN: Efficient Tampering Detection\\ of Deployed Neural Nets}

\date{}

\author{
  Erwan {Le Merrer}\\
     Univ Rennes, Inria, CNRS, Irisa\\
       erwan.le-merrer@inria.fr 
  \and Gilles Tr\'edan\\
       LAAS/CNRS\\
       gtredan@laas.fr
}

\maketitle

\begin{abstract}
  Neural networks are powering the deployment of embedded devices and Internet of Things. Applications range from personal assistants to critical ones such as self-driving cars. It has been shown recently that models obtained from neural nets can be \textit{trojaned}; an attacker can then trigger an arbitrary model behavior facing crafted inputs. This has a critical impact on the security and reliability of those deployed devices.

We introduce novel algorithms to detect the tampering with deployed models, \textit{classifiers} in particular. In the remote interaction setup we consider, the proposed strategy is to identify markers of the model input space that are likely to change class if the model is attacked, allowing a user to detect a possible tampering. This setup makes our proposal compatible with a wide range of scenarios, such as embedded models, or models exposed through prediction APIs.
We experiment those tampering detection algorithms  on the canonical MNIST dataset, over
three different types of neural nets, and facing five different
attacks (trojaning, quantization, fine-tuning, compression and
watermarking). We then validate over five large models (VGG16, VGG19,
ResNet, MobileNet, DenseNet) with a state of the art dataset
(VGGFace2), and report results demonstrating the possibility of an
efficient detection of model tampering.

\end{abstract}

\section{Introduction}

Neural network-based models 
are increasingly
embedded into systems that take autonomous decisions in place of
persons, such as in self-driving cars
\cite{Wu2017SqueezeDetUS,deepxplore} or in robots \cite{7965912}.
The value of those embedded models is then not only due to the 
investments for research and development, but also because of their
critical interaction with their environment. First attacks on
neural-based classifiers \cite{BIGGIO2018317} aimed at subverting
their predictions, by crafting inputs that are yet still perceived by
humans as unmodified. This leads to questions about the security of
applications embedding them in the real world
\cite{Kurakin2016AdversarialEI}, and more generally initiated the
interest of the security community for those problems
\cite{DBLP:journals/corr/PapernotMSW16}. Those \textit{adversarial}
attacks are confined in modifying the data inputs to send to a neural
model. Yet, very recently, other types of attacks were shown to
operate by \textit{modifying the model itself}, by embedding
information in the model weight matrices. This is the case of new
\textit{watermarking} techniques, that aim at embedding watermarks
into the model, in order to prove model
ownership\cite{Nagai2018,watermarking}, or of \textit{trojaning}
attacks \cite{trojannn,stuxdnn} that empowers the attacker to trigger
specific model behaviors.

\begin{figure}[h!]
  \centering
    \includegraphics[width=0.47\textwidth]{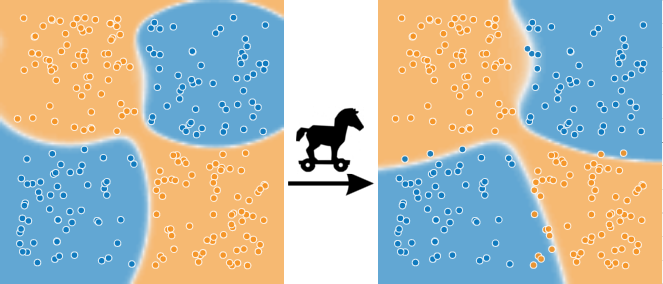}
  \caption{Illustration of a model tampering and its impact on the decision boundaries. We use the TensorFlow playground application for illustration; this toy dataset is fitted by a three hidden layer neural network (blue or yellow data inputs must be covered by their own color, ideally). The weights of solely the first layer have been modified (by removing $-0.1$ from them), to simulate an attack on the model. We observe slight changes of decision boundaries, that lead to misclassify some input data. This paper leverages such misclassifications for detecting the tampering attacks.\vspace{-0.4cm}}
  \label{example}
\end{figure}

The fact that neural network weights are by themselves implicit (\ie
they do not convey explicit behavior upon inspection as opposed to the
source code or formal specification of a software component) expose
applications to novel attack surfaces. It is obvious that those
attacks will have increasingly important impact in the future, and
that the model creators -- such as companies -- have a tremendous
interest in preventing the tampering with their models, specially in
embedded applications \cite{deepxplore,8123567}.

In this context, we are interested in providing a first approach to
detect the tampering of a neural network-based model that is deployed
on a device by the model creator. We stress that one can consider as
an attack \textit{any} action on a model (\ie on its weights)
that was not performed by the creator or the operator of that model.
To illustrate the practical effect of the tampering with a model, we
present in Figure \ref{example} a scenario of a fitted model (leftmost
image) over two classes of inputs (blue and yellow dots). After a
slight manipulation of few weights of the model (attack described in
the Figure caption), we can observe the resulting changes in the model
decision boundaries on the rightmost image. The attack caused a
movement of boundaries, that had the consequence at inference time to
return some erroneous predictions (\eg blue inputs in the yellow area
are now predicted as part of the wrong class). Those wrong predictions
might cause safety issues for the end-user, and are thus due to the
attack on the originally deployed model.

\subsection{Practical illustration of an attack: a trojaned classifier}

We motivate our work through a practical attack, using the technique
proposed by Liu et al. in \cite{trojannn}.  The goal of the attack is
to inject a malicious behaviour into an original model, that was already
trained and was functional for its original purpose.  The technique
consists in generating a \textit{trojan trigger}, and then to retrain
the model with synthesized datasets in order to embed that trigger
into the model. The trigger is designed to be added to a benign input,
such as an image, so that this new input triggers a classification
into the class that was targeted by the attacker. This results in a
biased classification triggered on demand, with potentially important
security consequences.  The model is then made available as an online
service or placed onto a compromised device; its behavior facing user
requests appears legitimate. The attacker triggers the trojan at will
by sending to the model an input containing the trojan trigger.

A face recognition model, known as the VGG Face model \cite{Parkhi15},
has been trojaned with the technique in \cite{trojannn} and made
available for
download \cite{troj} 
by the
authors of the attack.  We access both this model and its original
version.
The observed modifications in the weights do not indicate, a priori,
to which extent the accuracy has been modified (as depicted in Figure
\ref{example} where blue inputs get the orange label and vice-versa), as neural-based models have
highly implicit behaviours. One then has to pass a test dataset
through the classification API to assess the accuracy change.  The
authors report an average degradation of only $2.35\%$ over the
original test data, for the VGG Face trojaned model. In other words,
both models exhibit a highly similar behaviour (\ie similar
classifications) when facing benign inputs.
As a countermeasure to distinguish both models, an approach is compute
such a variation or, in a more straightforward approach, to compute
hashes of their weights (or hashes of the memory zone where the model
is mapped) and to compare them \cite{1027797}. However, this approach
requires a direct access to the model.

Unfortunately, models are not always (easily) accessible, because of
their embedding on IoT like devices for instance, as reported in work
by Roux et al. \cite{8123567}. This accessibility problem arises in at
least in two contexts of growing popularity: $i)$ if the model is
directly embedded on a device and $ii)$ if the model is in production on a
remote machine and only exposed through an API. In such cases where
the model is only accessible through its query API, one can think
about a black-box testing approach \cite{10.1007/3-540-54967-6_57}: it
queries the suspected model with specific inputs, and then compares
the outputs (the obtained classification labels) with those produced
by the original model.

In this paper, we study the efficiency of novel black-box testing
approaches. Indeed, while in theory testing the whole model input
space would for sure reveal any effective tampering, this would
require an impractical amount of requests.  For this approach to be
practical, the inputs used for querying the model shall be chosen
wisely. Our core proposal is then to identify specific inputs whose
property is to change classification as soon as the model is tampered
with. We refer to these crafted inputs as \textit{markers}.  Our
research question is: \textit{How to craft sets of markers for
  efficient model tampering detection, in a black-box setup?}

\subsection{Considered setup: black-box interactions}

While there exists a wide range of techniques, at the system or hardware level, to harden the tampering with software stacks on a device (please refer to the Related Work Section), the case of neural network models is salient.
Due to the intrinsic accuracy change of the model accuracy after the attack (cf Figure \ref{example}), we argue that stealthy, lightweight, and application-level algorithms can be designed in order to detect attacks on a remotely executed model.

 We consider the setup where a \textit{challenger}, possibly a company
 that has deployed the model on its devices, queries a \textit{remote}
 device with standard queries for classifying objects. This is
 possible through the standard classification API that the device
 exposes for its operation (paper \cite{stealing} discusses the same
 query setup for classification APIs of web-services).  To a query
 with a given input object (\eg image, sound file) is answered
 (inferred) a class label among the set of classes the model has been
 trained for; this is the nominal operation of a neural
 classifier. Note that for maximal applicability, we do not assume
 that the device is returning probability vectors along with classes;
 such an assumption would make the problem easier, but also less
 widely applicable (as it is known from previous attacks for stealing
 models that those probabilities should \textit{not} be returned, for security hardening
 facing attackers attempts \cite{stealing,7958568}).


\subsubsection{Contributions}
The contributions of this paper are:\\
\textit{(i)} to introduce the problem of the tampering detection of neural
      networks in the back-box setup, and to formalize this problem in
      Section \ref{sec:bb}.\\
\textit{(ii)} to propose three algorithms to address the
      challenge of efficient detection, by crafting markers that serve
      as attack witnesses. Those are compared to a strawman
      approach. Each have their own scope and efficiency on certain types of neural
      networks facing different attacks.\\
\textit{(iii)} to extensively experiment those algorithms on the canonical
      MNIST dataset, trained on three state of the art neural network
      architectures. We do not only consider the attack of trojaning,
      but also generalize to any attempt to modify the weights of the
      remote neural network, through actions such as fine-tuning,
      compression, quantization, and watermarking.  We then validate
      the efficiency of our approach over five large public models,
      designed for image classification (VGG16, VGG19, ResNet,
      MobileNet and DenseNet).

\subsubsection{Organization of the paper}
The remaining of this paper is organized as follows. We first precisely define the black-box interaction setup in the context of classification, and define the tampering detection problem in that regard in Section \ref{sec:bb}. The algorithms we propose are presented in Section \ref{s:algos}, before they are experimented with in Section \ref{s:experiments}; the limits of the black-box setup for tampering detection are also presented in that Section.
We finally review the Related Work in Section \ref{s:related} and conclude in Section \ref{s:conclusion}.


\section{Design rationale}
\label{sec:bb}
\subsection{The black-box model observation setup}
\label{ss:bb}

We study neural network classifiers, that account for the largest
portion of neural network models. Let $d$ be the dimensionality of the
input (feature) space $X$, and $C$ the set of target labels of
cardinality $n=|C|$. Let $\M:\mathbb{R}^d \to C$ be a classifier
model for the problem\footnote{By \textit{model}, we mean the trained
  model from a deep neural network architecture, along with its
  hyper-parameters and resulting weights.}, that takes inputs $x\in X$
to produce a label $\hat{y} \in C$: $\mathcal M(x) = \hat{y}$.

To precisely define the notion of decision boundary in this context we need the posterior probability vector estimated by $\M$:
$\{ \mathbb{P}(c|x,\M), c\in C\}$. When the context is clear we will
omit $x$ and $\M$ to simply write $\mathbb{P}(c)$. Note that while internally $\M$
needs to generate this vector to produce its estimate $\hat{y} =
\argmax_{c\in C} (\mathbb{P}(c))$, we assume a  black-box observation where only
$\hat{y}$ is available to the user.

\begin{definition}[Black-box model]
The challenger queries the observed model $\M$ with arbitrary inputs $x
\in X$, and gets in return $\M(x) = \hat{y} \in C$.
\label{blackbox}
\end{definition}
This setup is strictly more difficult than a popular setup where the
probability vector is made available to the user (but that can lead to misuses, as shown in
 in \cite{stealing} by ``stealing'' the remote model).

We now give a definition of the decision boundary of a model. This definition is adapted from
\cite{554193} for the black-box setup.
 
\begin{definition}[Decision boundary]
Given $\M$, an input $x \in X$ is on the decision boundary between
 classes if there exists at least two classes $c_i,c_j\in C$
maximising the posterior probability: $\mathbb{P}(c_i)=\mathbb{P}(c_j)=\max_{c\in C} \mathbb{P}(c)$. 

The set of points on decision boundaries defines a partitioning of
input space $X$ into $C$ equivalence classes, each class containing
inputs for which $\M$ predicts the same label.
\label{boundary_point}
\end{definition}

We can now provide the definition of models distinguishability when
queried in a black-box setup, based on the returned labels.

\begin{definition}[Models distinguishability]
  Two models  $\M$ and $\M'\neq \M$ can be distinguished in a black-box
  context if and only if
$\exists~x \in X~s.t.~\M(x) \ne \M'(x).$
\label{indistinguishable}
\end{definition}
Note that this implies that their decision boundaries are not equivalent
on input space $X$. Conversely, two different 
classifiers that end up through a training configuration to raise the
same decision boundaries on input space $X$, with equivalent classes
in each region, are \textit{indistinguishable} from an observer in the
black-box model.

Indistinguishability is trivially the inability to distinguish two models.
Since it is generally impossible to test the whole input set $X$, a
restricted but practical variant is to consider the
indistinguishability of two models with regards to a given set of inputs,
$\alpha$.  We name this variant \textit{$\alpha$-set
  indistinguishability}.

Since we are interested in cases where $\M'$ is a slight alteration of
the original model $\M$, it is interesting to quantify their
differences over a given set $S$:

\begin{definition}[Difference between models] Given two models $\M$,
  $\M'$ and a set $S$, the difference between $\M$ and
  $\M'$ on $S$ is defined as:
$\Delta(\M,\M', S)=\frac{1}{|S|}.\sum_{x \in S} \delta(\M(x),\M'(x)),$ with $\delta(.)=1$ if $\M(x) \ne \M'(x)$, $0$ otherwise.
\label{distance}
\end{definition}
In this light, two $S$-set indistinguishable models $\M$ and $\M'$ have by definition $\Delta(\M,\M',S)=0$.

\begin{definition}[Model stability]
Given an observation interval in time $\mathcal{I}=[t_0,t_e]$, and a model $\M$ observed at time $t_0$ (noted $\M_{t_0}$), the
model $\M$ is stable if it is indistinguishable from $\M_{t_0} \forall t \in \mathcal{I} \setminus t_0$.
\label{stability}
\end{definition}

\subsection{The problem of tampering detection}

\begin{problem}[Tampering detection of a model in the black-box setup]
Detect that model is not stable, according to Definition \ref{stability}.
\label{problem}
\end{problem}

This means that finding one input $x$ so that $\M_{t_0}(x) \ne \M_{t'}(x)$
is sufficient to solve Problem 1. 
Consequently, an \textit{optimal}
algorithm for solving Problem \ref{problem} is an algorithm that
provides, for any model $\mathcal{M}$ to defend, a \textit{single}
input that is guaranteed to see its classification changed on the
attacked model $\mathcal{M}'$, for any possible attack.

Since it is very unlikely, due to the current understanding of neural
networks, to find such an optimal algorithm, we resort to finding
sets of markers (specific data inputs) whose likelihood to change
classification is high as soon as the model is tampered with.
Since the challenge is to be often executed, and be as stealthy as possible, we refine the research question:
\textit{Are there algorithms to produce sensitive and small marker sets allowing model tampering detection in the black-box setup?}
More formally, we seek algorithms that given a model $\M$ find sets of inputs $K \subset X$
of low cost and high sensitivity to attacks:

\begin{definition}[Tampering detection metrics]
~\begin{itemize}
\item \textit{Cost:} $\vert K \vert$, the number of requests
  needed to evaluate\\ $\{\mathcal{M}(x), \forall x \in K \}$.
\item \textit{Sensitivity:} The probability to detect any model tampering, namely that at least one input (marker) of $K$ gets a different label. Formally, the performance of a set $K$ can be defined as $\mathbb{P}( \exists x \in K $ s.t. $ \mathcal{M}(x)\neq \mathcal{M'}(x), \forall \mathcal{M'})$.
\end{itemize}
\end{definition}

It is easy to see that both metrics are bound by a tradeoff: the bigger the marker set, the more likely it contains a marker that will change if the model is tampered with. We now have a closer look to this relation.

\subsubsection{Cost-sensitivity tradeoff}

As stated in the previous Section, we assume in this paper that the remote
model returns the minimal information, by building our algorithms with
solely labels being returned as answers ($\M(x) \in C$).

Let $\M$ an original model, and $\M'$ a tampered version of
$\M$. Consider a set of inputs $K\subset X$ of cardinality (cost)
$s$. Let $p=\mathbb{P}(\M(x)\neq \M'(x))$ be the probability that a
marker $x \in K$ triggers (that is, $x$ allows the detection of the
tampering). In the experiments, we refer to estimations of $p$ as the
\textit{marker triggering ratio}.  Assume that given a model tampering
the probabilities of each marker to trigger are independent. The
overall probability to  detect the tampering by querying
with $K$ of size $s$ is the sensitivity $c= 1-(1-p)^s$. While the
challenger is tempted to make this probability as high as possible,
he also desires to keep $s$ low to limit the cost of such an
operation, and to remain stealthy.

\begin{figure}[h!]
\vspace{-0.3cm}
  \centering
    \includegraphics[width=0.4\textwidth]{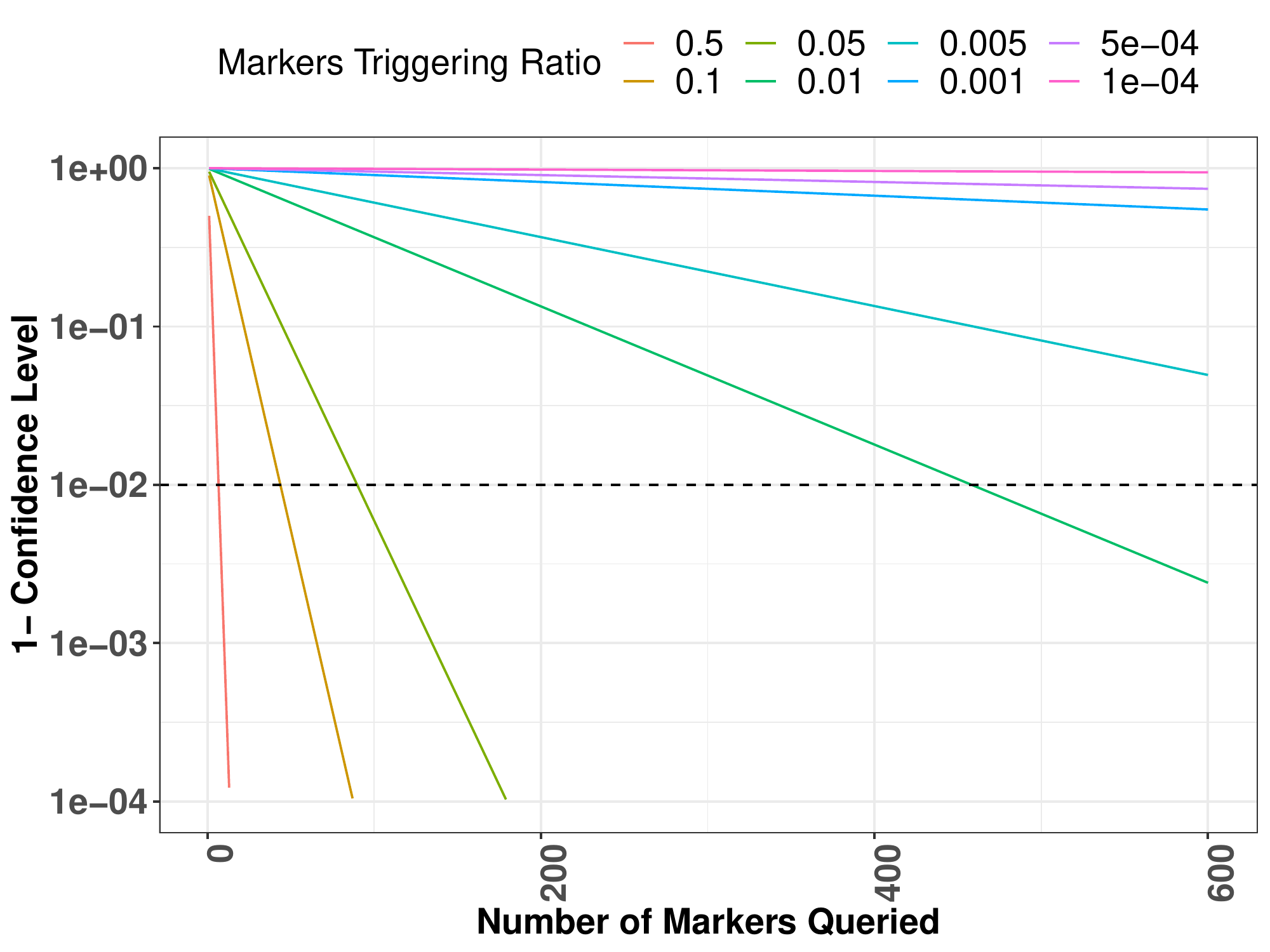}
  \caption{$y$-axis: Inverse Confidence Level for (\ie probability of
    failing at)  detecting a tampered model, with the amount of queries on the $x$-axis, and for a given marker triggering ratio (indicated on the top legend).\vspace{-0.2cm}}
    \label{VGG-pval}
\end{figure}

In general, we assume the challenger first fixes the desired
sensitivity  $c$ as a confidence level of his decisions (say for instance $99\%$ confidence, $c=.99$). It turns out that one can easily derive the minimum key size $s$ given $p$ and $c$:

\begin{align*}
   (1-p)^s <  1-c \\
   s \times log(1-p)  <  log(1-c)\\
   s  >  log(1-c)/log(1-p) 
\end{align*}
These relations highlight the importance of having a high marker
triggering ratio: there is an exponential relation between $p$ and
$s$. This relation is illustrated Figure~\ref{VGG-pval}, that relates
the key size $s$, the marker triggering ratio, and the chosen
confidence. Please note the inverted logscale on the $y$-axis. It
shows that for a constant confidence $c=.99$ (dashed line), a key
composed by markers easily triggered will be small (\eg 6 for
$p=0.5$), while a key composed by markers with a low trigger ratio
will be considerably larger (458 for $p=0.01$ for instance).

This short analysis is stressing the importance of designing efficient algorithms that craft markers which have a high chance to be misclassified after the attack; this permits to challenge the model with less markers.

Finally, let us note that the goal of an attacker, beside
  the successful implementation of his attack (\eg the embedding of the
  trojan trigger in the method proposed in \cite{trojannn}), is to
  have the minimal impact on the model accuracy for obvious
  stealthiness reasons (\eg the claim for only $2.35\%$ degradation in
  \cite{trojannn}). We thus have the classical conflicting interests
  for the attacker and the challenger: the challenger hopes for a
  noticeable attack in order to detect it more easily, while the
  attacker leans towards the perfect attack leaving no trace at all.
  This motivates the research for the crafting of very sensitive
  markers, in order to detect even the slightest attacks on a model.
 
\section{Algorithms for the Detection of Remote Model Tampering}
\label{s:algos}

\subsection{Reasoning facing black-box model classifications}

Since we aim at leveraging a limited set of marker queries 
to challenge the model, and because success probability is involved,
let us introduce notions of true and false positives for
Problem 1. A true positive refers to a tampered model being detected
as such, a true negative corresponds to the legitimate (original)
model being recognized as original.

\subsubsection{False negatives in tampering detection} Regarding the problem of tampering detection in the black-box setup, a false negative for the detection occurs if an algorithm challenges the model, gets classifications, and concludes that the model has not been tampered with, while it was actually the case. The probability of failure that we presented in Figure \ref{VGG-pval} thus constitutes the probability of false negatives.

\subsubsection{False positives in tampering detection} In this setup,
assuming that the original model has a deterministic behaviour, false
positives cannot happen. A false positive would be the detection of an
attack that did not occur. Since the challenger had full control over
his model before deployment, he knows the original labels of his
markers. Any change in those labels is by the problem definition an
indication of the tampering and therefore cannot be misinterpreted as an
attack that never occurred.

\subsection{The strawman approach and three algorithms}

The purpose of the novel algorithms we are now presenting is to
provide the challenger with a key $K$ to query the remote model, that
is composed of a set of $|K|=s$ input markers. Concretely, given an
original model $\M$, those algorithms generate a set of markers
$K$. The owner of the model stores $K$ along with the response vector $\hat Y_\M$,
associated with the markers in $K$:
$\hat Y_\M \gets \mathcal{M}(i), \forall i \in K.$

We assume that $K$ is kept secret (unknown to the attacker). When the user needs to
challenge a potentially attacked model $\M'$, it simply queries $\M'$
with inputs in $K$, and compares the obtained response vector
$\hat Y_{\M'}$ with the stored $\hat Y_{\M}$. If the two response
vectors differ, the user can conclude that $\M\neq \M'$.

\begin{table*}[]\footnotesize
\centering
  \begin{tabular}{c|c|c|c|c|c}
Remote Opacity   & \multicolumn{4}{c}{\cellcolor[HTML]{DAE8FC}Black-Box}                                                & \cellcolor[HTML]{FE996B}White-Box \\
Original Opacity & \multicolumn{2}{c}{\cellcolor[HTML]{DAE8FC}Black-Box}                 & \multicolumn{3}{c}{\cellcolor[HTML]{FE996B}White-Box}                                                            \\
Knowledge Required     & Input format & A bunch of inputs & Original weights & Original architecture & Remote weights              \\
Algorithm              & GRID                      & SM                & WGHT                   & BADV                        & --                               
\end{tabular}
\caption{Summary of Algorithm requirements, ordered by the amount of
  knowledge required for the challenge.\vspace{-0.5cm}}
\label{tabalg}
\end{table*}

The algorithms are designed to operate on widely different aspects of
black-box model querying. Table~\ref{tabalg} synthesizes the different
knowledge exploited by those algorithms. Cases where the remote model
is accessible (\ie white-box testing) can be solved by direct weight
comparison or standard software integrity testing approaches
\cite{1027797}, and are not in the scope of this paper.

\begin{figure}[h!]
  \centering

\subfigure[][]{%
\label{fig:ex3-a}%
\includegraphics[height=1.1in]{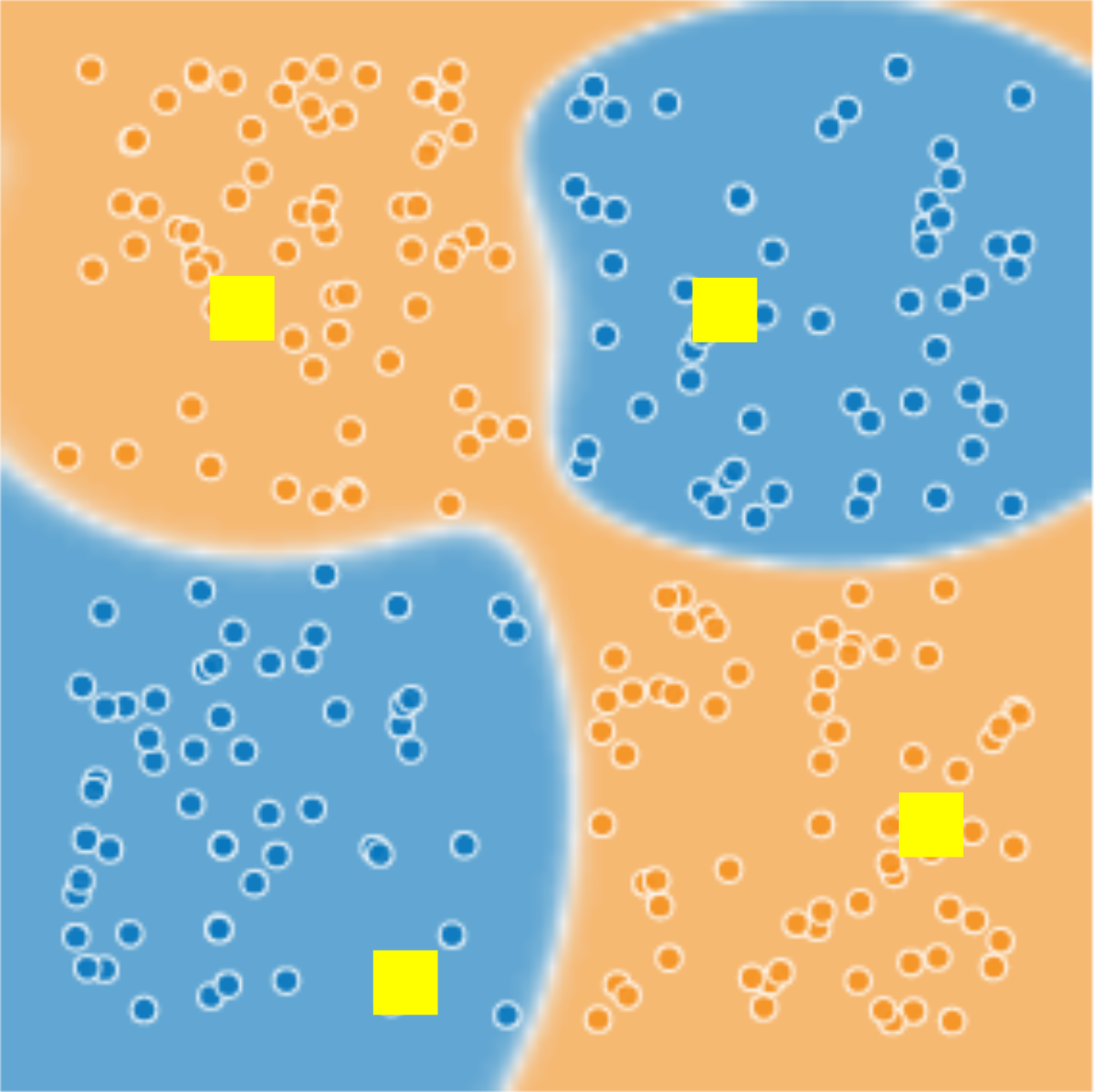}}%
\hspace{8pt}%
\subfigure[][]{%
\label{fig:ex3-b}%
\includegraphics[height=1.1in]{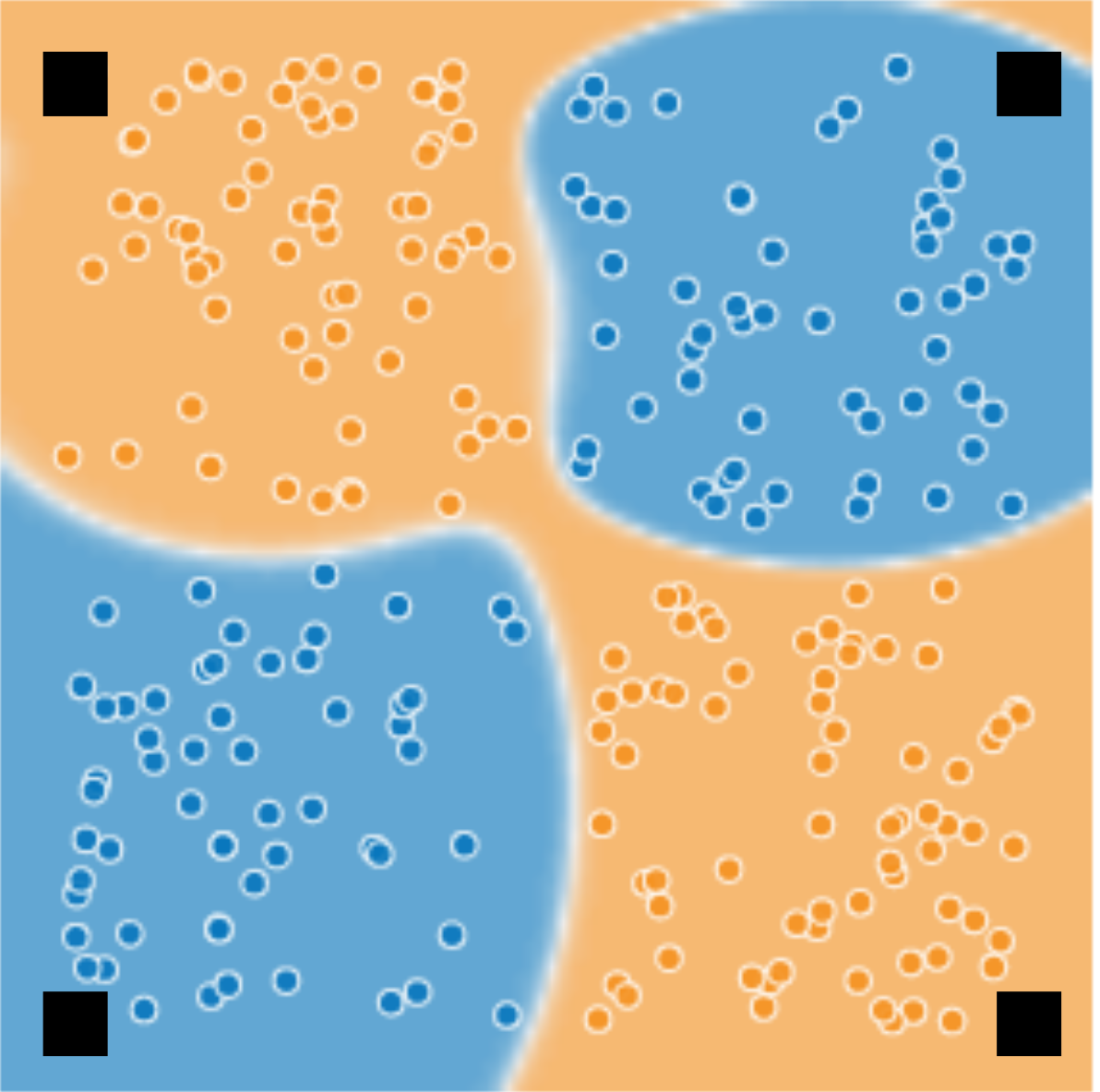}} \\
\vspace{-0.25cm}
\subfigure[][]{%
\label{fig:ex3-c}%
\includegraphics[height=1.1in]{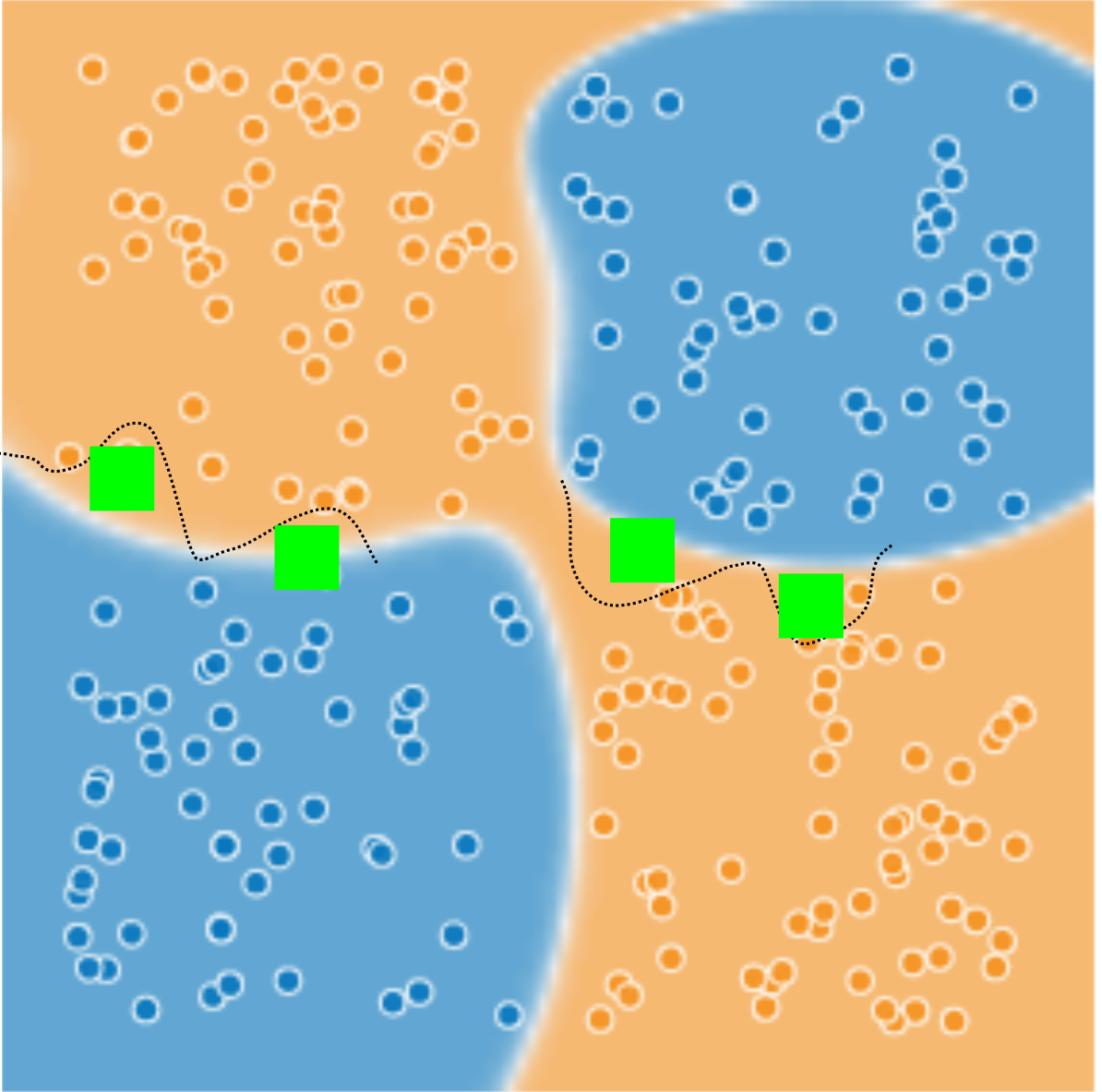}}%
\hspace{8pt}%
\subfigure[][]{%
\label{fig:ex3-d}%
\includegraphics[height=1.1in]{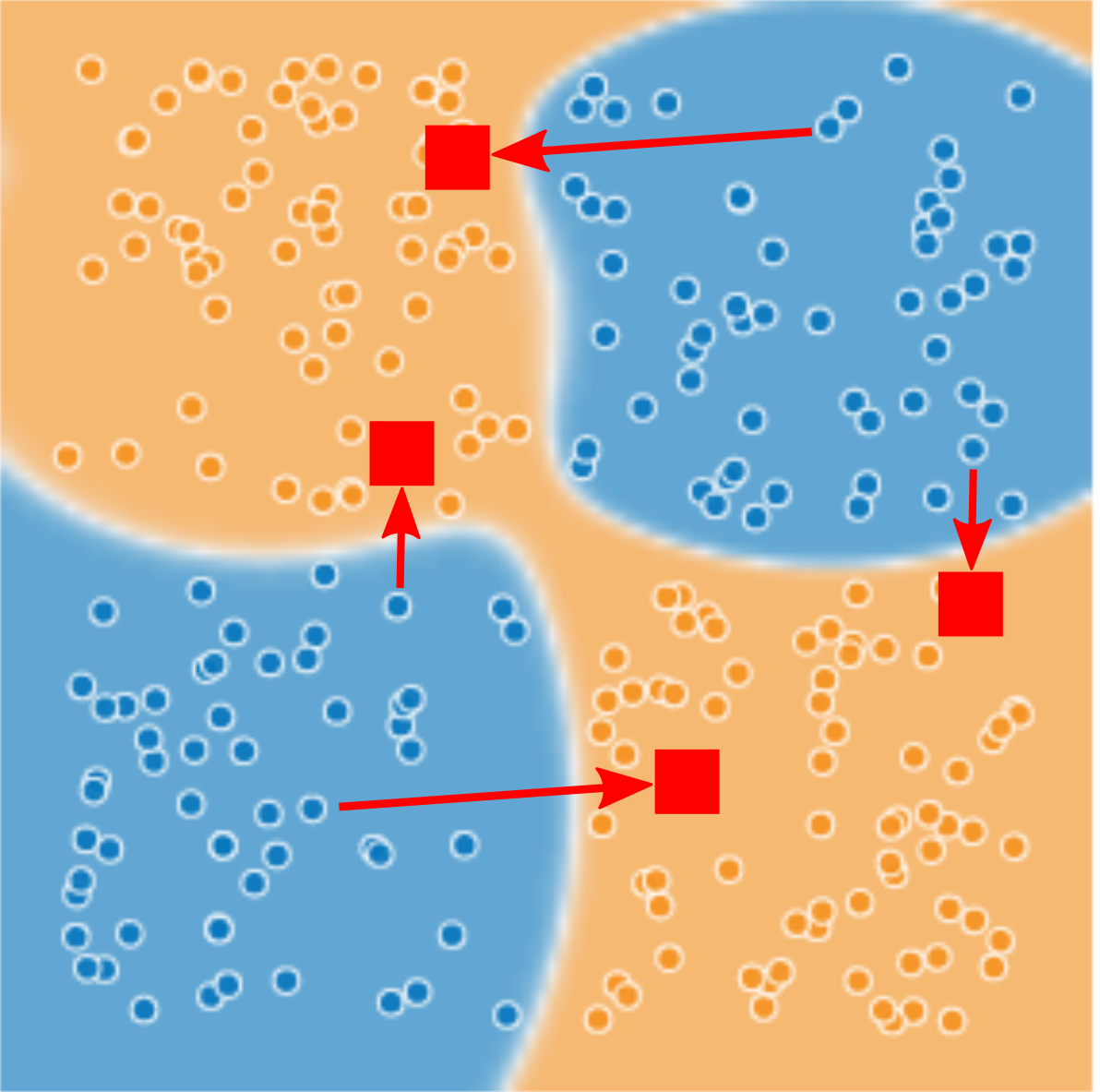}}%
\vspace{-0.3cm}
  \caption{The placement of markers by the four algorithms. \sm (a) picks a fraction of the test set inputs as markers; their position are thus related to the position of the dataset inputs. \gr (b) places markers at random corners of the $d$ dimensional hyper-cube delimiting the input space. \wg (c) finds inputs of the test set that are sensitive to weight perturbations, to select them as markers. Finally, \adv (d) converts test set inputs into adversaries located nearby the boundaries.\vspace{-0.3cm}}
  \label{markers}
\end{figure}

\sm (standing for \textit{strawman})
represents an intuitive approach, consisting in tracking a
sort of "non-regression" with regards to the initial model deployed on
devices. The \gr algorithm (\textit{grid-like inputs}) is also model
agnostic, as it generates inputs at random, that are expected to be
distant from real-data distribution, and then to assess boundary
changes in an efficient way. Both \wg (\textit{perturbation of
  weights}) and \adv (\textit{boundary adversarial examples}) take as
arguments the model to consider, and a value $\epsilon$; the former
applies a random perturbation on every weight to observe which are the most
sensitive inputs with regards to misclassification after the attack
(in the hope that those inputs will also be sensitive to forthcoming
attacks). The latter generates \textit{adversarial} inputs by the
decision boundaries, in the hope to be sensitive to their
movements. We now give their individual rationale in the following
subsections, and present in Figure \ref{markers} an illustration.

\subsubsection{The strawman approach (\sm)}
\begin{algorithm}
\KwIn{A test set $\mathcal{T}$; $s$}
\For{$0$ \textbf{to} $s$}{
   $K.append(\mathcal{T}[$random\_int$(|\mathcal{T}|)]))$ 
  }
\Return K
\caption{\sm}\label{alg.sm}
\end{algorithm}

This strawman approach uses inputs from the test set as markers in
order to assess changes in classification. An initial classification
is performed for the markers before model deployment, and the resulting classes are expected
to remain identical in subsequent queries.

Algorithm \ref{alg.sm} then simply returns a key of size $s$, from the random selection of inputs in the provided test set $\mathcal{T}$.

\subsubsection{Grid-like inputs (\gr)}
\begin{algorithm}
{\small
  \KwIn{Image width $x$ and height $y$; $s$}
\For{$0$ \textbf{to} $s$}{
    \For{$0$ \textbf{to} $x$}{
        \For{$0$ \textbf{to} $y$}{
           img$[x][y]\gets$ random\_bit$()$ 
           }
        }
   $K.append($img$)$
  }
\Return K
\caption{\gr}\label{alg.gr}
}
\end{algorithm}

This algorithm generates markers independently of the model $\mathcal{M}$ that is to be challenged.

Most generally, in classification tasks, the inputs are normalized
before their usage to train the model. Without loss of generality for
a normalization out of the $[0,1]$ range, for each of the $d$ dimensions
of the considered model (\eg the $784$ dimensions of a MNIST image),
Algorithm \ref{alg.gr} sets a random bit.  The rationale is to
generate markers that are far apart the actual probability
distribution of the base dataset: since the training and tampering
with $\mathcal{M}$ are willing to preserve accuracy, constraints are
placed on minimizing test set misclassification. The consequence is a
large degree of freedom for decision boundaries that are far apart the
mass of inputs from the training set. We thus expect those crafted
inputs to be very sensitive to the movement of boundaries resulting
from the attack.

\subsubsection{Perturbation of weights (\wg)}
~\\
\begin{algorithm}[h!]
{\small
  \KwIn{A test set $\mathcal{T}$; a model $\mathcal{M}$; a small $\epsilon$; $s$}
\For{$i \gets 0$ \textbf{to} $|\mathcal{T}|$}{
    pre$.append(\mathcal{M}(i))$ 
}

\For{$i \gets 0$ \textbf{to} $|get\_weights(\mathcal{M})|$}{
    $\mathcal{M}.set\_weight(i,\mathcal{M}.get\_weight(i)+ random\_float(-\epsilon,+\epsilon))$
    }
    
\For{$i \gets 0$ \textbf{to} $|\mathcal{T}|$}{
    post$.append(\mathcal{M}(i))$ 
}
\For{$i \gets 0$ \textbf{to} $s$}{
\uIf{pre($i$) $\ne$ post($i$)}{
    $K.append($pre($i$))
    }
    \uIf{$|K|=s$}{break;}
}
\tcc{Assumes $|K| = s$, increase $\epsilon$ otherwise}
\Return K
\caption{\wg}\label{alg.wg}
}
\end{algorithm}

The \wg algorithm takes as arguments the model $\mathcal{M}$, and a value $\epsilon$. It observes the classifications of inputs in dataset $\mathcal{T}$, before and after a perturbation has been applied to all weights of $\mathcal{M}$ (\ie a random perturbation of every weight to up to $\pm \epsilon$).
Inputs for which label have changed, due to this form of tampering, are sampled to populate key $K$.
The rationale is that with a low $\epsilon$, the key markers are expected to be very sensitive to the tampering of model $\mathcal{M}$.
In other words, inputs from $K$ are expected to be the most sensitive inputs from $\mathcal{T}$ when it comes to tamper with the weights of $\mathcal{M}$.

\subsubsection{Boundary adversarial examples (\adv)}
\begin{algorithm}
{\small
  \KwIn{A test set $\mathcal{T}$; a model $\mathcal{M}$; an attack $\mathcal{A}$; a small $\epsilon$; $s$}
\For{$i \gets 0$ \textbf{to} $|\mathcal{T}|$}{
    adv$.append(\mathcal{A}(\mathcal{M}(i),\epsilon))$
    
    \uIf{$\mathcal{M}(i) \ne \mathcal{M}(adv)$}{
    $K.append($adv$)$
    }
    \uIf{$|K|=s$}{break;}
}

\tcc{Assumes $|K| = s$, increase $\epsilon$ otherwise}
}
\caption{\adv}\label{alg.adv}
\end{algorithm}
Adversarial examples have been introduced in the early works presented in
\cite{BIGGIO2018317} and re-framed in
\cite{DBLP:journals/corr/SzegedyZSBEGF13}, in order to fool a
classifier (by making it misclassify inputs) solely due to slight
modifications of the inputs. Goodfellow et al. then proposed
\cite{Goodfellow:2015} an attack for applying perturbations to inputs
that leads to vast misclassifications of the provided inputs (that attack is named the \textit{fast gradient sign
  attack or FGSM}). Those crafted inputs yet appear very similar to
the original ones to humans, which leads to important security
concerns \cite{deepxplore}; note that since then, many other attacks
of that form were proposed (even based on different setup assumptions
\cite{Papernot:2017:PBA:3052973.3053009}), as well as platforms to
generate them (\eg \cite{clev} or \cite{fool}). 

We propose with \adv to leverage the FGSM attack, but in an adapted
way.  The FGSM attack adds the following quantity to a legitimate
input $x$: $\epsilon \times sign (\triangledown_x
J(\mathcal{M},x,y)),$ with $\triangledown_x$ being the gradient of $J$
(the cost function used to train model $\mathcal{M}$), and $y$ the
label of input $x$.  $\epsilon$ captures the intensity of the attack
on the input. Approach in \cite{Goodfellow:2015} is interested in
choosing an $\epsilon$ that is large enough so that most of the inputs in the
batch provided to the FGSM algorithm are misclassified (\eg
$\epsilon=0.25$ leads to the misclassification of $97.5\%$ of the
MNIST test set). We are instead interested in choosing an $\epsilon$
that is sufficient to create $s$ misclassified markers only; the
rationale is that the lower the $\epsilon$, the closer the crafted
inputs are to the decision boundary; our hypothesis is that this
proximity will make those inputs very sensitive to any attack of the
model that will even slightly modify the position of decision
boundaries.  In practice, and with Algorithm \ref{alg.adv}, we start
from a low $\epsilon$, and increase it until we get the desired key
length $s$.


\section{Experimental Evaluation}
\label{s:experiments}

This section is structured as follows: we first describe the
experiments on MNIST (along with the considered attacks and parameters
for algorithms). We then discuss and experiment the limitations of the
black-box setup we considered. We finally validate our take-aways on
five large image classification models, in the last subsection of this
evaluation.

We conduct experiments using the TensorFlow platform, 
using the Keras library.

\subsection{Three neural networks for the MNIST dataset}
The dataset used for those extensive experiments is the canonical MNIST database of handwritten digits, that consists of $60,000$ images as the training set, and of $10,000$ for the test set. The purpose of the neural networks we trained are of classifying images into one of the ten classes of the dataset. 

The three off-the-shelf neural network architectures we use are available on the Keras website \cite{keras}, 
namely as {\tt mnist\_mlp} (0.984\% accuracy at 10 epochs), {\tt mnist\_cnn} (0.993\% at 10) and {\tt mnist\_irnn} (0\
.9918\% at 900). We rename those into MLP, CNN and IRNN respectively.
They constitute different characteristic architectures, with one historical multi-layer \textit{perceptron}, a network with \textit{convolutional} layers, and finally a network with \textit{recurrent} units.

\begin{figure}[t!]
  \centering
    \includegraphics[width=0.51\textwidth]{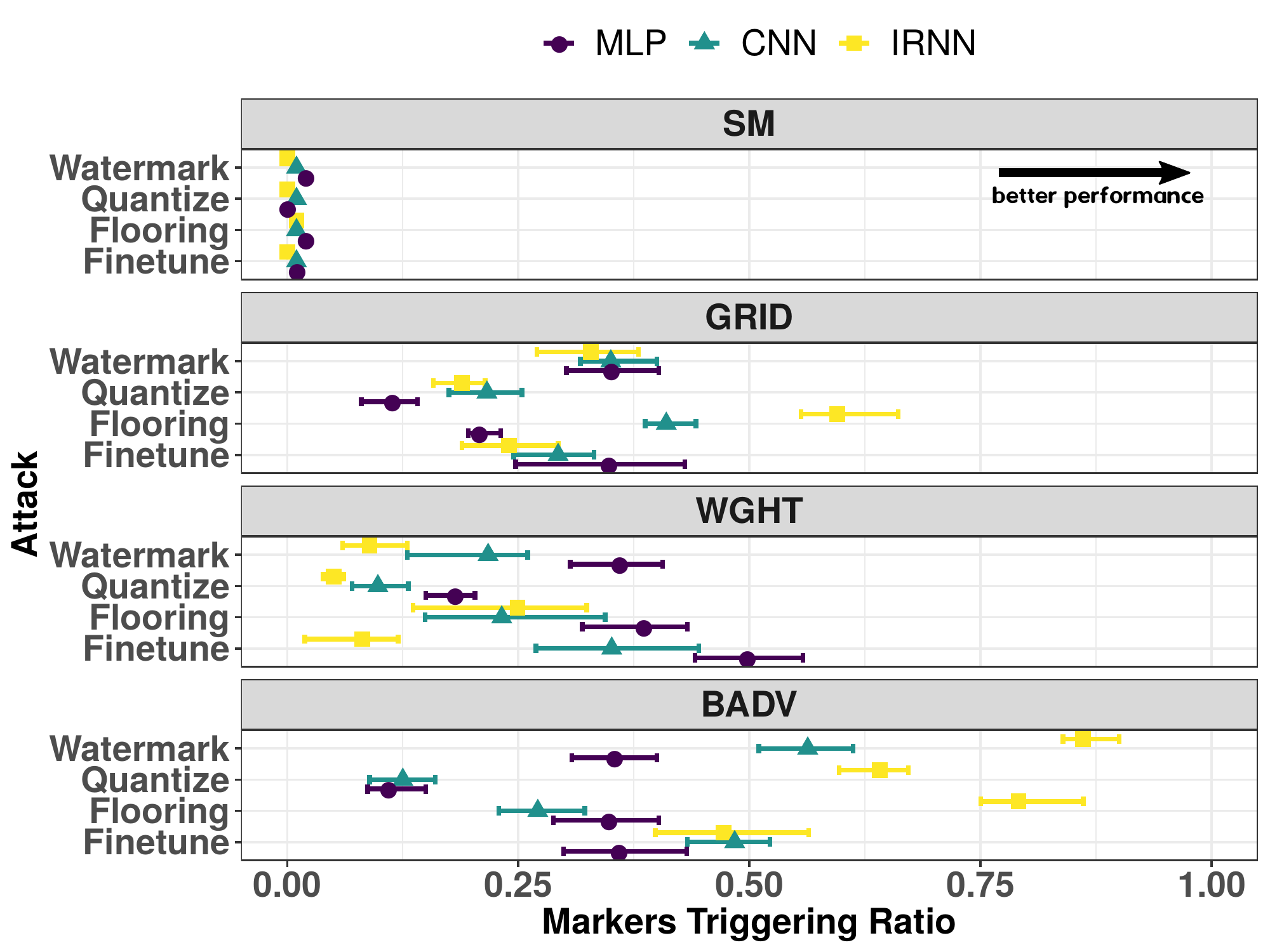}
  \caption{Performance results of the proposed algorithms (\gr, \wg and \adv), with regards to the strawman approach (\sm). Attacks applied to the models (listed on top) are indicated on the $y$-axis, while the ratio of triggered markers is indicated on the $x$-axis. Average results, as well as standard deviations for the algorithms are presented, the rightmost the better for their performance. For instance the \wg algorithm, facing the fine-tuning attack, sees half of its key markers being triggered, while only one marker (ratio of $1/100$) is triggered for the strawman approach.\vspace{-0.4cm}}
\label{fig:results}
\end{figure}

\subsection{Attacks: from quantization to trojaning}

This subsection lists the five attacks we considered. Excluding the watermarking and trojaning attacks, the others are standard operations over trained models; yet if an operator has already deployed its models on devices, any of those can be considered as attacks, as they tamper with the model that was designed for a precise purpose.

\subsubsection{Quantization attack}
This operation aims at reducing the number of bits representing
each weight in the trained model. It is in practice widely
used prior to deployment in order to fit the architecture and
constraints of the target device. TensorFlow by default uses 32-bit
floating points, and the goal is convert the model into 8-bit integers
for instance.  The TensorFlow 
\texttt{fake\_quant\_with\_min\_max\_args~}~ function
is used to simulate the quantization of the
trained neural network. We kept the default parameters of that
function (8-bits quantization, with -6 to 6 as clamping range for the
input).

\subsubsection{Compression attack}
A form of compression is \textit{flooring}; it consists in setting to zero all model weights that are below a fixed threshold, and aims at saving storage space for the model.
We set the following threshold value for the three networks: 0.0727, 0.050735 and 0.341 for the MLP, CNN and IRNN networks respectively.
Those thresholds cause the degradation of network accuracies by about one percent (accuracies after the compression are 0.9749, 0.9829 and 0.9821, respectively).

\subsubsection{Fine-tuning attack}
Its consists in starting from a trained model, and to re-train it over
a small batch of new data. This results in model weight changes, as
the model was adapted through back-propagation to prediction errors
made on that batch. We used a random sample of $300$ inputs from the
MNIST test set for that purpose.

\subsubsection{Watermarking attack}
Watermarking techniques \cite{Nagai2018,watermarking} embed
information into the target model weights in order to mark its
provenance.  Since work in \cite{watermarking} operated on the MNIST
dataset, and provided detailed parameters, we implemented this
watermarking technique on the same models (MLP, CNN, and IRNN). The
watermark insertion proceeds by fine-tuning the model over adversarial
examples to re-integrate them into their original class, in order to
obtain specific classifications for specific input queries (thus
constituting a watermark). This approach requires a parameter for the
mark that we set to $0.1$, consistently with remarks made in
\cite{watermarking} for maintaining the watermarked model accuracy.

\subsubsection{Trojaning attack}
We leverage the code provided in a GitHub repository, under the name
of Stux-DNN, and that aims at trojaning a convolutional neural network
for the MNIST dataset \cite{stuxdnn}.  We first train the provided
original model, and obtain an accuracy of $93.97\%$ over the MNIST test
set.  The trojaning is also achieved with the provided code.

\medskip
After applying those five attacks, the models accuracies changed;
those are summarized on Table \ref{degradations}. Note that some attacks may
surprisingly result in a slight accuracy improvement, as this is the
case for MLP and quantization.

\begin{table}
\center
\resizebox{0.4\textwidth}{!}{
{\small
   \hspace{-0.3cm}\begin{tabular}{|c|ccc|}
    \hline
    ~ & MLP & CNN (Stux) & IRNN \\ \hline
Original model accuracy  &  0.9849   & 0.9932 (0.9397\textdagger)  & 0.9919   \\ \rowcolor{Gray}
Quantization  &  0.9851  & 0.9928  &  0.9916 \\  \rowcolor{Gray}
Flooring  & 0.9749    & 0.9829   & 0.9821 \\  \rowcolor{Gray}
Fine-tuning  & 0.9754 & 0.9799  & 0.9917  \\  \rowcolor{Gray}
Watermarking \cite{watermarking} & 0.9748 & 0.9886  & 0.9915  \\  \rowcolor{Gray}
Trojaning L0 \cite{stuxdnn}  & - &   (0.9340\textdagger) & -    \\  \rowcolor{Gray}
Trojaning mask \cite{stuxdnn}  & - &  (0.9369\textdagger)   & - \\  
   \hline
\end{tabular}
   }
}
\caption{Original model accuraries (white row), and accuracies
  resulting from attacks (grey rows). The lower the loss in accuracy,
  the stealthier the attack. Values marked \textdagger~are obtained on
the trojaned CNN model introduced in \cite{stuxdnn}, and publicly available
on the authors website.\vspace{-0.5cm}}
\label{degradations}
\end{table}

\subsection{Algorithms settings}

\subsubsection{Settings for \sm} \sm uses a sample of images from the original MNIST test set, selected at random.

\subsubsection{Settings for \gr} We use the Python Numpy uniform random generator for populating markers, that are images of 28x28 pixels.

\subsubsection{Settings for \wg} All the weights in the model are perturbed by adding to each of them a random float within $[-0.07,+0.07]$, $[-0.07,+0.07]$ or $[-0.245,+0.245]$ for the MLP, CNN and IRNN architectures respectively. This operation must keep the accuracy loss within a small percentage, while making it possible to cause enough classification changes for populating $K$ (those values allowed to identify just over $100$ markers).

\subsubsection{Settings for \adv} For generating adversarial examples that are part of the key, we leverage the Cleverhans Python library \cite{clev}. 
The FGSM algorithm used in \adv, requires the $\epsilon$ parameter for the perturbation of inputs to \textit{(i)} be small enough, and \textit{(ii)} allow for the generation at least $200$ adversarial examples out of $10,000$ files in the test set. $\epsilon$ is set to $0.04$, $0.08$ and $0.14$ for the MLP, CNN and IRNN networks. 

\subsection{Experimental results}

Results are presented in Figure \ref{fig:results}, for all the attacks
(excluding the trojaning attack), the three models and the four
algorithms. We set key size $s=100$; each experiment is executed $10$
times.

\sm generates markers that trigger with a  probability below 0.02 for all attacks and all models; this means that some attacks such as for instance quantization over the MLP or IRNN models remain undetected after $100$ query challenges.

All three proposed algorithms significantly beat that strawman approach; the most efficient algorithm, on average and in relative top-performances is \adv. Most notably, on the IRNN model it manages to trigger a ratio of up to 0.791 of markers, that is around $80\%$ of them, for the flooring attack. This validates the intuition that creating sensitive markers from adversarial examples by the boundary (\ie with small $\epsilon$ values) is possible.

The third observation is that \gr arrives in second position for general performances: this simple algorithm, that operates solely on the data input space for generating markers, manages to be sensitive to boundary movements.

The \wg algorithm has high performance peaks for the MLP model, with up to half of triggered markers for the fine-tuning attack, and a ratio of $0.385$ for flooring (\ie more than one third of markers are triggered); it has the lowest performances of the three proposed algorithms, specifically for the IRNN model. This may come from the functioning of its recurrent architecture that makes it more robust to direct perturbations of weights: the model is more stable during learning (it requires around 900 epochs to be trained, while the two other models need only 10 epochs to reach their peak accuracy).

The watermark attack is very well detected on the IRNN model with the \adv algorithm (ratio of $0.86$), on an equivalent rate on three models by \gr, while \sm still shows trace amount of markers triggered for MLP and CNN, and none for IRNN.

Considering the relatively low degradation of the models reported on Table \ref{degradations} (\ie within around $1\%$ maximum)\footnote{Trojaning attacks in \cite{trojannn} reports degradation over the original models of $2.60\%$ (VGG Face recognition), $3\%$ (speech recognition) or $3.50\%$ (speech altitude recognition).}, we conclude that all three proposed algorithms capture efficiently the effects of even small attacks on the models to be defended, while \sm would only be valuable in cases of large degradation of models. We illustrate in the subsection \ref{ss:undetectable} the degradation-detectability trade-off.

\begin{figure}[h!]
  \centering
    \includegraphics[width=0.35\textwidth]{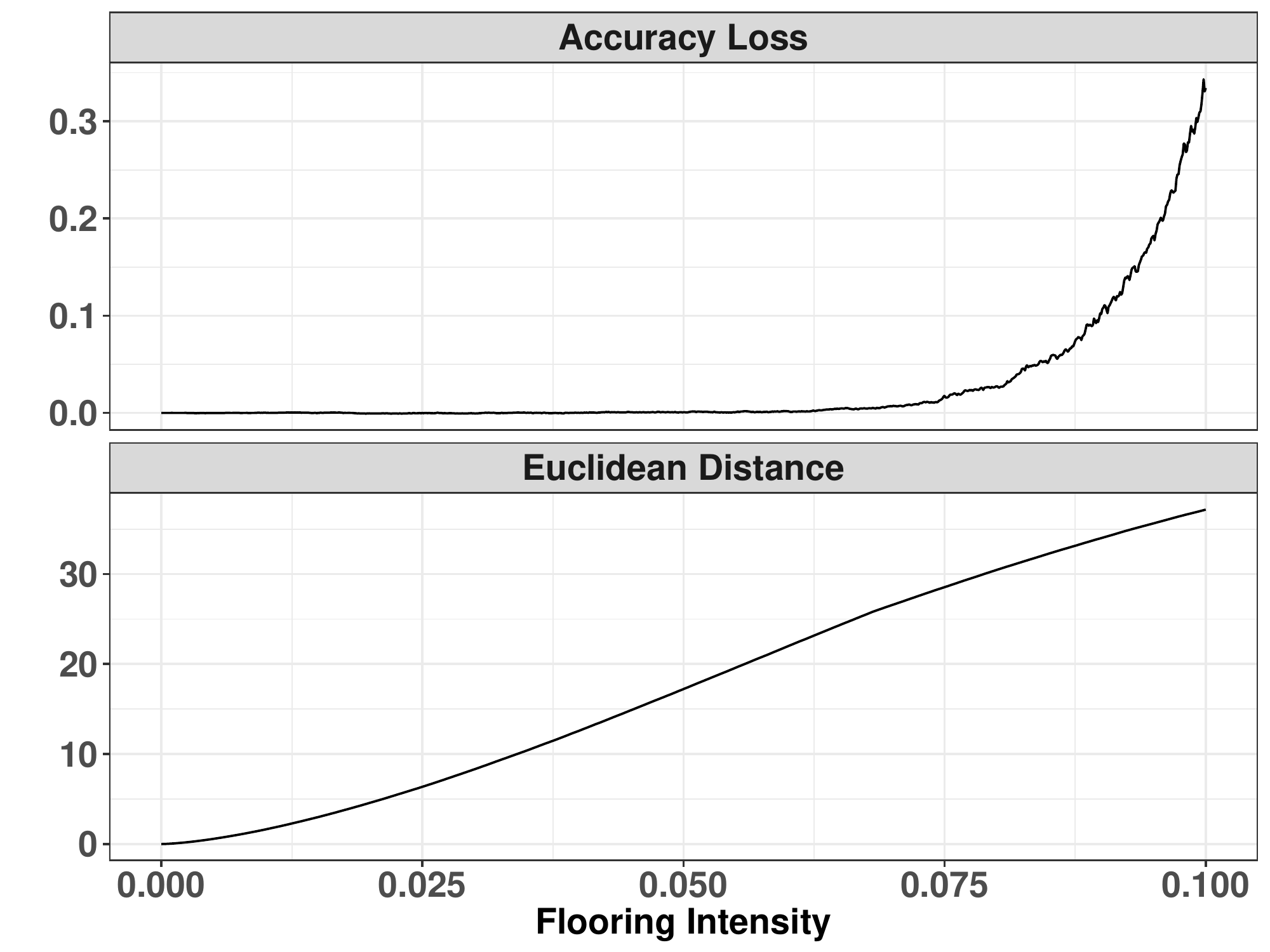}
  \caption{Applying a compression (flooring) attack of increasing intensity ($x$-axis). Top-Figure: loss in accuracy of the attacked model, as compared to the original one. Bottom-Figure: measure of the Euclidean distance between the weights of the original and attacked models.\vspace{-0.45cm}}
\label{fig:inc_attack}
\end{figure}

\begin{figure}[h!]
  \centering
    \includegraphics[width=0.35\textwidth]{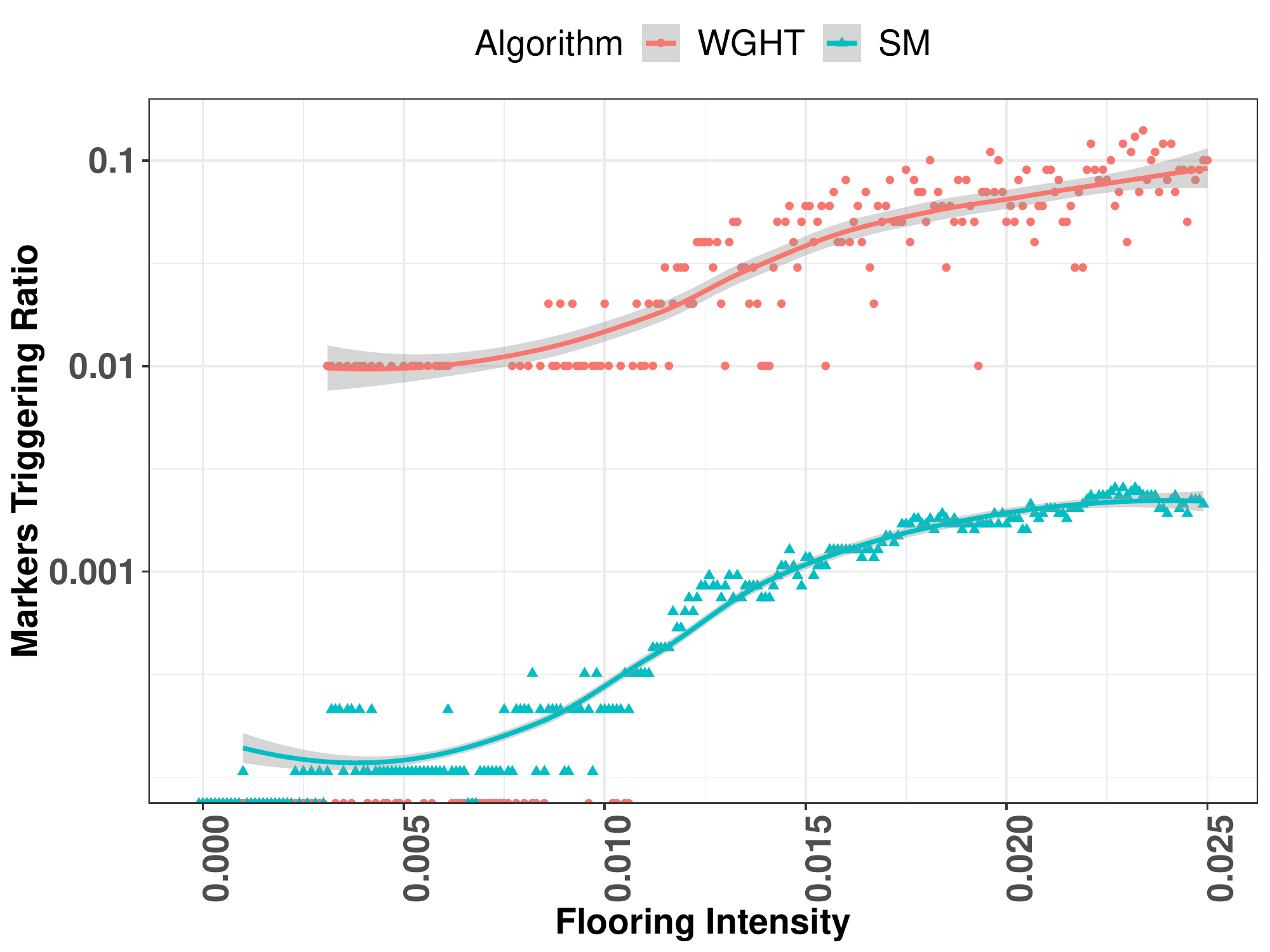}
  \caption{Impact of the same progressive flooring attack than in
    Figure \ref{fig:inc_attack} on the ratio of markers triggered for
    both \sm and \wg ($y$-axis, logscale).\vspace{-0.9cm}}
\label{fig:inc_markers}
\end{figure}

\subsection{Validation on a trojaning attack on MNIST}

The modified accuracy of the neural network model proposed
by \cite{stuxdnn}, due to the attack, is reported on Table \ref{degradations}. The attack has two trojaning modes (L0
and mask). 
We now question the ability
of one of our algorithms to also be outperforming the strawman
approach \sm; we experiment with \gr.  Results are that \sm manages to trigger ratios of
$0.0524$ and $0.0529$ of markers, for L0 and mask modes respectively
(please refer to their original paper for details on these
techniques).  \gr reaches ratios of $0.4560$ and $0.4502$, that are
8.7x and 8.5x increases in efficiency. This suggests that for a practical
usage, a small key $K$ of
$s=10$ will detect the attack, while \sm is likely to raise a false
negative.

\subsection{Undetectable attacks and indistinguishably: illustrations}
\label{ss:undetectable}

We now present examples that illustrate the inherent limits of a black-box interaction setup, as defined in subsection \ref{ss:bb}.

Let's consider the MLP model, along its best performing algorithm for
tamper detection, \wg. Assume that the model is tampered with using
compression (flooring), and that we observe successive attacks: from
an attack value (flooring threshold $v$) starting at 0, it reaches a value of $v=0.1$ by
increments of $0.001$ (\ie at each attack, weights under $v$ are
floored). We observe the results after every attack; we plot in Figure
\ref{fig:inc_attack} the loss in accuracy of the attacked model
(top-Figure), and the Euclidean distance between the original and attacked model weights
(bottom-Figure). For instance, we observe a $30\%$ accuracy loss at a
distance of 40.  Since the loss is noticeable from around $v=0.05$, we
zoom in for plotting the corresponding ratio of markers triggered in
Figure \ref{fig:inc_markers}.

Those two figures convey two different observations, presented in the
next two subsections.

\subsubsection{Limits of the algorithms and of the black-box setup}

The following cases may happen for attacks that have a very small impact on the model weights. 
\begin{itemize}
    \item \textit{Case 1:} Accuracy changed after the attack, but the algorithm failed in finding at least one input that has changed class (\ie no marker from $K$ has shown a classification change).
    \item \textit{Case 2:} We did not manage to find any such input despite the attack.
\end{itemize}

\textit{Case 1} for instance occurs with $v=0.004$, as seen in Figure \ref{fig:inc_markers}. This means that both algorithms have failed, for the chosen key length $s$, to provide markers that were present in the zones where boundary moved due to the attack. (Please remind that, if the accuracy post attack has been modified, this means that some inputs from the test set has a changed label, then indicating that boundaries have de facto moved).

\textit{Case 2} is particularly interesting as it permits to illustrate Definition \ref{indistinguishable} in its restricted form: an attack occurred on $\mathcal{M}$ (as witness by a positive Euclidean distance between the two models in Figure \ref{fig:inc_attack}), but it does not result in a measurable accuracy change. It is $\alpha$-set indistinguishable, with here $\alpha$ being the MNIST test set. We measure this case for $v=0.003$, where pre and post accuracies are both 0.9849.

\textit{Case 1} motivates the proposal of new algorithms for the problem. We nevertheless highlight that the trojan attack \cite{trojannn} degrades the model on a basis of around $2\%$, while our algorithm is here unable to detect a tampering that is two order of magnitude smaller (accuracy loss of $0.02\%$ for the attacked model). This indicate extreme cases for all future tampering detection approaches. \textit{Case 2} questions the black-box interaction setup. This setup enables tampering detection in a lightweight and stealthy fashion, but may cause indecision due to the inability to conclude on tampering due to the lack of test data that can assess accuracy changes. 

\subsubsection{\wg outperforms \sm by nearly two orders of magnitude for small attacks}
As observed in Figure \ref{fig:inc_markers}, the \sm markers triggered ratio ranges from $0.0001$ to around $0.005$, while for \wg it ranges from $0.01$ to $0.1$, in this extreme case for attack detection with very low model degradation. 

\begin{figure}[h!]
  \centering
    \includegraphics[width=0.45\textwidth]{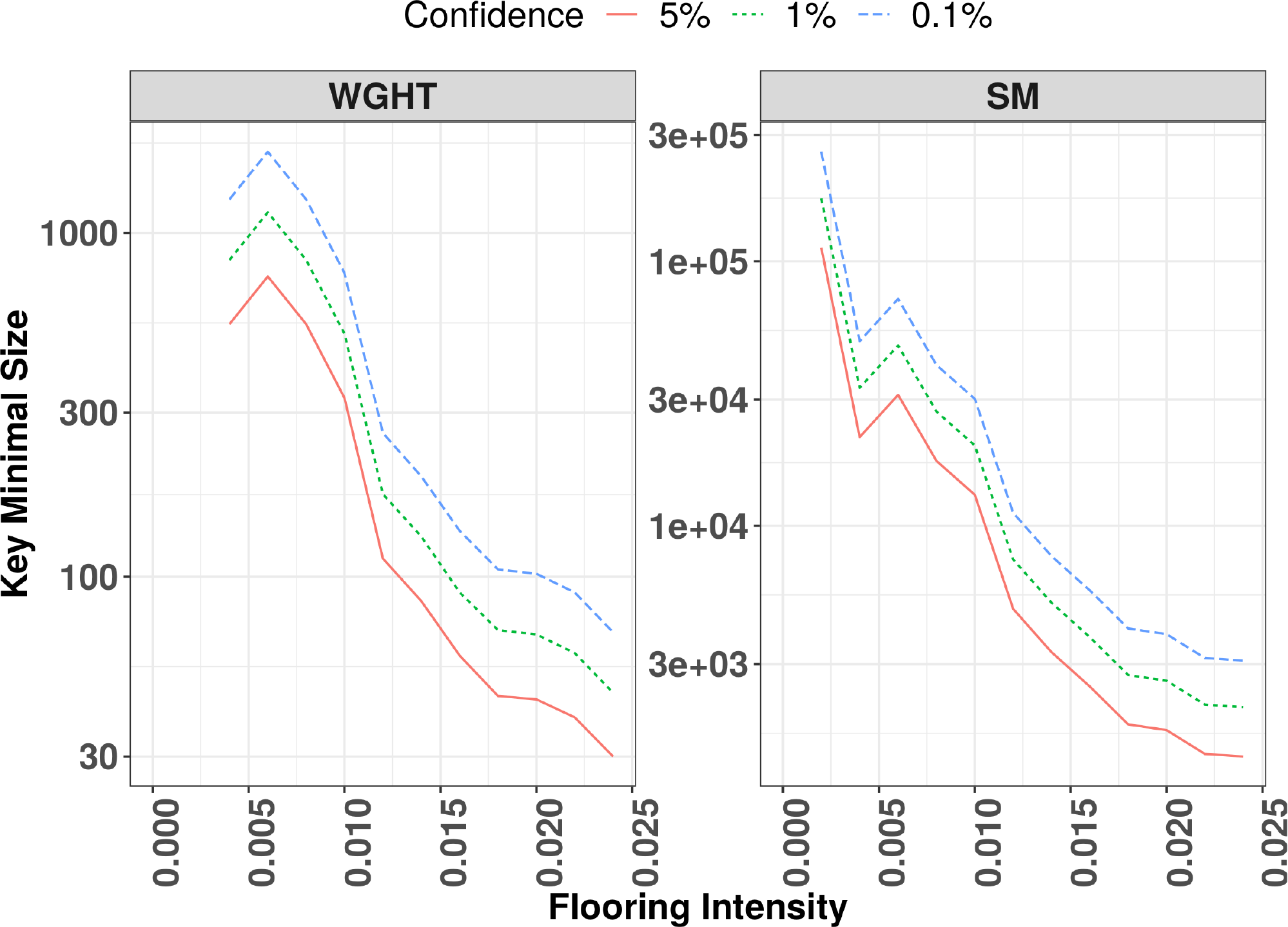}
  \caption{Key $K$ size $s$ ($y$-axis) to choose for a given challenge
    detection failure (noted confidence, on top-legend).  Lines
    represent a smoothed average.  Intuitively, the smaller the
    failure probability, the larger the key to select; this is the
    experimental counterpart to the analysis in Figure
    \ref{VGG-pval}. Its size depends on the chosen algorithm and on the intensity of the flooring attack
    ($x$-axis).\vspace{-0.4cm}}
\label{fig:inc_size}
\end{figure}

Figure \ref{fig:inc_size} concludes this experiment by presenting the key size $s$ that is to be chosen, depending on the algorithm and on the tolerance to attack intensity. This is in direct relation with the efficiency gap observed on previous figure: the more efficient the algorithm for finding sensitive markers, the smaller the query key for satisfying the according detection confidence.
For an equivalent confidence, the key size to leverage for \sm is $100$ times longer than for the \wg algorithm, confirming the efficiency of the techniques we proposed in this paper.

\subsection{Validation on five large classifier models}

We conducted extensive experiments on the standard MNIST dataset for
it allows computations to run in a reasonable amount of time, due to
the limited sizes of both its datasets and of the models for learning
it. In order to validate the general claim of this paper, we now
perform experiments on five large and recent models for image
classification, using a state of the art dataset.
This validation is interested in checking the consistency with the
observation from the MNIST experiments, that have shown that our
algorithms significantly outperform the strawman approach. 

We leverage five open-sourced and pre-trained models:
VGG16\cite{VGG16} (containing 138,357,544 parameters, as compared to MNIST
  models containing 669,706 (MLP), 710,218 (CNN) and 199,434 (IRNN)
  parameters), VGG19\cite{VGG16} (143,667,240 parameters),
ResNet50\cite{ResNet} (25,636,712 parameters),
MobileNet\cite{MobileNet} (4,253,864 parameters) and
DenseNet121\cite{DenseNet} (8,062,504 parameters).  Except for the two
VGG variants VGG16 and VGG19, all four architectures are broadly
different models, that each were proposed as independent improvements
for those image classification tasks (please refer to
  the Keras site \cite{keras} for each their own characteristics).

The VGGFace2 dataset has been made public recently \cite{Cao18}; it
consists of a split of training (8631 labels) and test (500
labels) sets. The labels in the two sets are disjoint.  We
consider a random sample of $10,000$ images of the VGGFace2 test
dataset, for serving as inputs to the \sm and \wg algorithms.  We note
that despite that labels in the test set are different from the
ones learnt in the models, this is a classic procedure (used \eg for
experiments in work by Liu et al. \cite{trojannn}): a neural network
with good performances will output stable features for new images, and
thus in our case predict consistently the same class for each new given
input. Those images are imported as 224x224 pixel images to query
the tested models (versus 28x28 for MNIST).
As for
previous experiment (Figure \ref{fig:inc_markers}), we experiment with
the \sm and \wg algorithms and $s=100$, with the flooring attack.  We perform the
computations on four Nvidia TESLAs V100 with 32 Gb of RAM each; each
setup is run three times and results are averaged (standard deviations
are presented).

\begin{figure*}[h]
  \centering
    \includegraphics[width=1.01\textwidth]{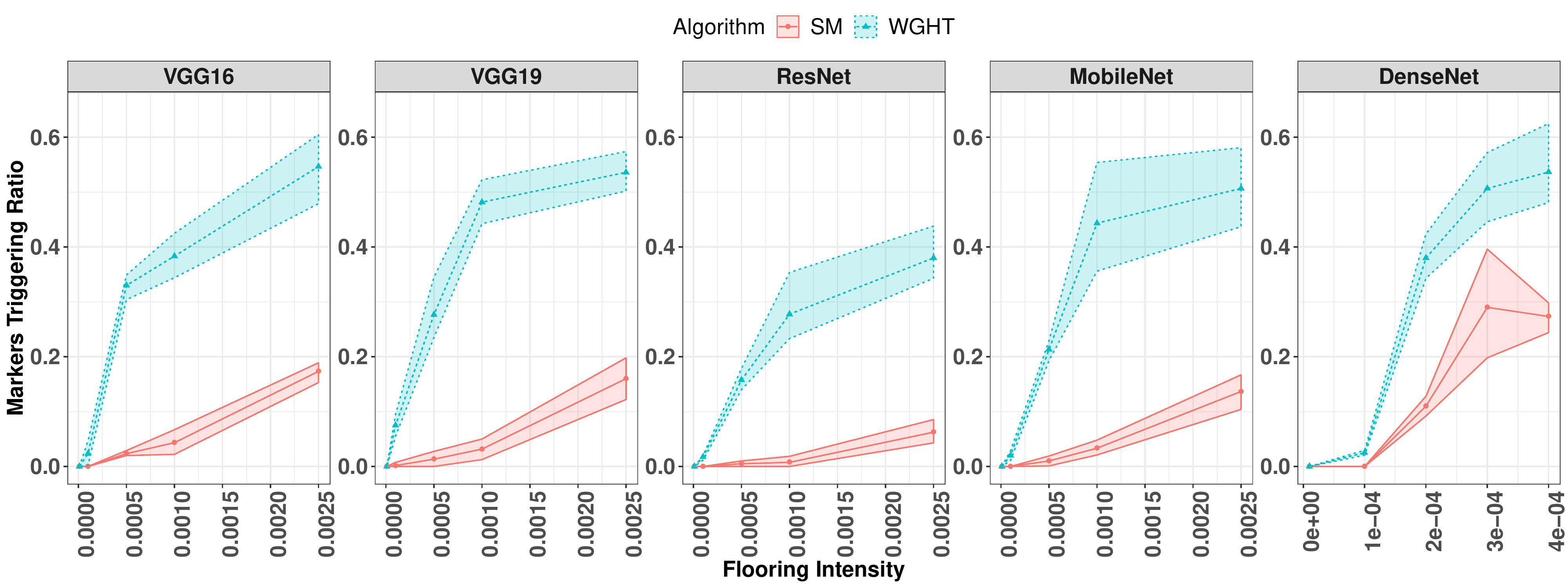}
\vspace{-0.5cm}
    \caption{Marker triggering ratio for five large image classification
  models. One proposed algorithm (\wg) versus the strawman approach
  (\sm), facing the flooring attack. A sample of the VGGFace2 test dataset is used.\vspace{-0.4cm}}
\label{fig:zoo}
\end{figure*}

Figure \ref{fig:zoo} presents the results. The $x$-axis of each figure
represents the flooring intensity,
  with the same values for all models, except for DenseNet
  because of its noticeable sensitivity to attacks.
VGG16 corresponds to the neural network architecture of trojaned in
paper \cite{trojannn}.  For all models, we observe that an attack of $0.00001$ is bellow
what both \sm and \wg can detect (situation presented in Section
\ref{ss:undetectable}). For the second smaller considered attack
values on the $x$-axis, only \wg manages to trigger markers; this
constitutes another evidence that crafted markers are more sensitive
and will trigger first for the smallest detectable attacks.  For all
the remaining flooring parameters, \sm triggers markers, but always
significantly less than \wg (up a factor of 15 times less, at
$x=0.001$ on VGG19). All the models exhibit a very similar trend for
both curves. The triggering ratio in the case of ResNet is lower for
both \wg and \sm, while gap between the two approaches remains
similar.  Finally, in the DenseNet case, we note a higher triggering
ratio for \sm than for other models on the last three flooring values;
the results are still largely in favor of the \wg algorithm.


\section{Related Work}
\label{s:related}

Research works targeting the security of embedded devices such
as IoT devices\cite{8123567}, suggest that traditional security
mechanisms must be reconsidered. Main problems for traditional
security tools is the difficulty to fix software flaws on those
devices, the frequency to which those flows are reported, and finally
their limited resources for the implementation of efficient protections.

\textit{Anti-tampering} techniques for traditional software applications may be applied directly on the host machine in some defense scenarios. This is the case for the direct examination of the suspected piece of software \cite{1027797}. 
\textit{Remote attestation} techniques \cite{Coker2011} allows for the distant checking of potential illegitimate modifications of software or hardware components; this is nevertheless requiring the deployment of a specific challenge/response framework (often using cryptographic schemes), that both parties should comply with.
\textit{Program result checking} (or black-box testing) is an old technique that inspects the compliance of a software by observing outputs on some inputs with expected results \cite{10.1007/3-540-54967-6_57}; it has been applied in conventional software applications, but not on the particular case of deep learning models, where the actions are driven by an interpretation of a model (its weights in particular) at runtime. In that light, the work we proposed in this paper is a form of result checking for neural model integrity attestation. Since it is intractable to iterate over all possible inputs of a neural network model to fully characterize it (unlike for reverse engineering finite state machines \cite{4400167} for instance), due to the dimensionality of inputs in current applications, the challenger is bound to create some algorithms to find some specific inputs that will carry the desired observations. 

After a fast improvement of the results provided by neural network-based learning techniques in the past years, models found practical deployments into user devices \cite{Lane:2016:DSA:2959355.2959378}. The domain of security for those models is a nascent field \cite{DBLP:journals/corr/PapernotMSW16}, following the discovery of several types of attacks. The first one is the intriguing properties of adversarial attacks \cite{BIGGIO2018317,Kurakin2016AdversarialEI,Goodfellow:2015,DBLP:journals/corr/SzegedyZSBEGF13} for fooling classifications; a wide range of proposals are attempting to circumvent those attacks \cite{7546524,Meng:2017:MTD:3133956.3134057,DBLP:journals/corr/XuEQ17}.
Counter measures for preventing the stealing of machine learning models such as neural networks thought prediction APIs are discussed in \cite{stealing}; it includes the recommendation for the service provider not to send probability vectors along with labels in online classification services. Some attacks are willing to leak information about individual data records that were used during the training of a model \cite{Song:2017:MLM:3133956.3134077,7958568}; countermeasures are to restrict the precision of probability vectors returned by the queries, or to limit those vectors solely to top-$k$ classes \cite{7958568}.
The possibility to embed information within the models themselves with watermarking techniques \cite{Nagai2018,watermarking} is being discussed on the watermark removal side by approaches like \cite{wat-removal}.
Trojaning attacks \cite{trojannn,stuxdnn} are yet not addressed, except by this paper, that introduced the problem and brought three novel algorithms to detect the tampering with models in a black-box setup.

\section{Conclusion}
\label{s:conclusion}

Neural network-based models enable applications to
reach new levels of quality of service for the end-user.  Those
outstanding performances are in balance, facing the risks that are
highlighted by new attacks issued by researchers and practitioners.
This paper introduced the problem of tampering detection for remotely
executed models, by the use of their standard API for
classification. We proposed algorithms that craft markers to query the
model with; the challenger detects an attack on the remote model by
observing prediction changes on those markers. We have shown a high level of
performance as compared to a strawman approach that would use inputs
from classic test sets for that purpose; the challenger can then
expect to detect a tampering with very few queries to the remote
model, avoiding false negatives. We believe that this
application-level security checks, that operate at the model level and
then at the granularity of the input data itself, is raising
interesting futureworks for the community.

While we experimented those algorithms facing small modifications made
to the model by attacks, we have also shown that below a certain level of
modification, the black-box setup may not permit to detect tampering
attacks. In other situations, where the attack is observed in practice
through accuracy change in the model, our algorithms can fail in the
detection task. Some even more sensitive approaches might be proposed
in the future. We believe this is an interesting futurework direction,
that is to be linked with the growing understanding of the inner
functioning of neural networks, and on their resilience facing
attacks.

\bibliographystyle{abbrv}
\bibliography{biblio}

\end{document}